\documentclass{aastex62}

\submitjournal{PASP}

\usepackage{chngcntr} 

\newcommand{\vdisp}{v$_{disp}$} 
\newcommand{\Ha}{EW(H$\alpha$)}

\newcommand{\Hd}{EW(H$\delta$)}
\newcommand{\Sersicamp}{Sersic\_amp\_R}
\newcommand{\Sersicn}{Sersic\_n\_R}
\newcommand{\sigmabalmer}{$\sigma_{\mathrm{balmer}}$}
\newcommand{\sigmaforb}{$\sigma\_{\mathrm{forb}}$}
\newcommand{\OII}{EW(OII$_{3729}$)}

\newcommand{\OIIIfour}{EW(OIII$_{4363}$)}
\newcommand{\OIIIfive}{EW(OIII$_{5007}$)}

\newcommand{\NII}{EW(NII$_{6584}$)}
\newcommand{\SII}{EW(SII$_{6731}$)}
\newcommand{\Hdabs}{H$\delta_{abs}$}
\newcommand{\Hbabs}{H$\beta_{abs}$} 
\newcommand{\Haabs}{H$\alpha_{abs}$}
\newcommand{\Hgabs}{H$\gamma_{abs}$}
\newcommand{\CaK}{Ca\_K$_{\mathrm{abs}}$}

\newcommand{\Sersicrzero}{Sersic\_r0\_R}
\newcommand{\Sersicrfifty}{Sersic\_r50\_R}
\newcommand{\Sersicrninety}{Sersic\_r90\_R}
\newcommand{\SersicC}{$\mathcal{C}$}

\accepted{Dec 10, 2018}

\shorttitle{Unsupervised Classification of Galaxies}
\shortauthors{Chattopadhyay et al.}

\begin{document}
\title{Unsupervised Classification of Galaxies. I. ICA feature selection}

\correspondingauthor{T. Chattopadhyay}
\email{tchatappmath@caluniv.ac.in}

\author{T. Chattopadhyay}
\affiliation{Department of Applied Mathematics, University of Calcutta, Kolkata, India }
\nocollaboration\noaffiliation

\author{D. Fraix-Burnet}
\affiliation{Univ. Grenoble Alpes, CNRS, IPAG, 38000 Grenoble, France}
\nocollaboration\noaffiliation\noaffiliation

\author{S. Mondal}
\affiliation{Department of Statistics, Bethune College, Kolkata, India}
\nocollaboration\noaffiliation

    \begin{abstract}
Subjective classification of galaxies can mislead us in the quest
        of the origin regarding formation and evolution of galaxies since this is necessarily limited to a few features.
        The human mind is not able to apprehend the complex correlations in a manyfold parameter space, and multivariate analyses are the best tools to
        understand the differences among various kinds of objects. In this series of papers, an objective classification
        of 362,923 galaxies from the Value Added Galaxy Catalogue (VAGC) is carried out
        with the help of two methods of multivariate analysis. First, Independent
        Component Analysis (ICA) is used to determine a set of derived independent
        components that are linear combinations of 47 observed features
        (viz. ionized lines, Lick indices, photometric and morphological properties,
        star formation rates etc.) of the galaxies. Subsequently, a
        K-means cluster analysis is applied on the nine independent components
        to obtain ten distinct and homogeneous groups.
In this first paper, we describe the methods and the main results. It appears that the nine Independent Components represent a complete physical description of galaxies (velocity dispersion, ionisation, metallicity, surface
brightness and structure). We find that our ten groups can be essentially placed into traditional and empirical classes (from colour-magnitude and emission-line diagnostic diagrams, early- vs late-types) despite the classical corresponding features (colour, line ratios and morphology) being not significantly correlated with the nine Independent Components.
More detailed physical interpretation of the groups will be performed in subsequent papers.
    \end{abstract}
    
    \keywords{galaxies: general -- galaxies: evolution -- methods: statistical}

    \section{Introduction}
    \label{introduction}

    Investigating the formation and evolution of galaxies is becoming a complicated process with the increased
    availability of huge databases as a result of instrumental improvements. A good understanding
    of the underlying physical processes requires synthetic numerical simulations of the formation and evolution of galaxies in an evolving Universe. 
    According to various studies \citep[e.g. see reviews in ][]{Silk2012,Genel2014}, classical formation of galaxies have been proposed
    to follow five major models: (i) the monolithic collapse model, (ii) the major merger model,
    (iii) the multiphase dissipational collapse model, (iv) accretion and (v) in situ hierarchical
    merging. But no model uniquely explains the formation of all galaxies.
    
    The historical and still most common approach to the classification of galaxies, is based on physical criteria, such as apparent features (i.e. morphology, emission line properties etc.) or more or less understood processes (starbursts, Active Galactic Nuclei (AGN) etc.). 
    Hubble's subjective
    classification based on galaxy morphology ignores many significant
    observable features such as kinematics, chemical composition etc. Physically based classifications also rely on only very few attributes among the numerous ones available: colour, size, environment, star formation rate, emission lines. All these classifications are certainly useful to study a specific property, but they cannot embrace the whole diversity and complexity of processes that shaped the galaxies.
 
    With the advent of multi-wavelength and multivariate databases, the goal of many studies has been to find the properties that best characterize the established classes. One such example is given by the Baldwin, Phillips \& Tarlevich (BPT) diagrams \citep{Baldwin1981,Veilleux1987} built from a few ratios of emission lines such as [OIII]/H$\alpha$, [OIII]/H$\beta$, [NII]/H$\beta$, [SII]/H$\alpha$ and [OI]/H$\alpha$. Delimitation lines have been defined empirically to separate different kinds of ionizing sources (AGNs, Low-Ionization Nuclear Emission-Line Regions (LINERs), Seyfert galaxies, star forming regions). But a multivariate analysis using [OIII]/H$\alpha$, [NII]/H$\beta$ and \Ha\ have shown that the empirical classes are not all statistically supported \citep{Souza2017}. One probable explanation is that several kinds of ionizing sources can be present in a single galaxy, so that a classification built from a two-feature diagram cannot reasonably translate the real complexity of even a single galaxy. 
    
    These classification approaches can be called object-based, to be opposed to data-driven science in which some characteristics are revealed by the accumulation of data and attributes. The so-called bimodality of galaxies \citep{Strateva2001,Eales2018} is an example of the latter: the observations of millions of galaxies have revealed that for several properties, galaxies show two distribution peaks that do not easily match pre-established physically motivated classes \citep[e.g.][]{Eales2018,Evans2018}. By chance, this bimodal structure can be seen with the eye on 2D diagrams.
  
  Extragalactic studies have now entered the statistical era challenging our approaches to understand the physics of galaxies and their evolution. 
  In most cases, there is no simple mathematical relationship between a given physical process and the relevant observable attributes. In addition, these attributes are generally multiple so that classifications based on only one or two features are not adequate to represent the physics of galaxies. Despite this caveat, many studies have embarked into complex statistical techniques (Bayesian tools, Deep Learning) to tackle the avalanche of data that astronomy is about to generate with the goal to obtain an automatic classification \citep[e.g.][]{Brescia2015}. However, these approaches, most often multivariate, use supervised learning techniques, i.e. the algorithm is trained on labelled data. The problem here is that the labels come from pre-established classifications based on very few features, in other words the algorithm is trained to find the results of human subjectivity \citep[e.g.][]{Cavuoti2014}. 
  
  In this context, one is tempted to apply statistical  unsupervised classification, such as a multivariate partitioning analysis to
    find homogeneous groups, not focusing on only one aspect of the physics of galaxies, but by exploring the cluster structure of the dataset \citep{Fraix-Burnet2015}. The goal is to use data-driven science to study the formation and evolutionary history of galaxies. One basic tool is Principal Component Analysis (PCA) and
    it has been used by many authors \citep[e.g.][]{Whitmore1984,Watanabe1985,Cabanac2002,Chattopadhyay2006,Peth2015}. However this is not a classification tool. It may be useful to represent pre-established classes using some of the Principal Components, but its main interest is a reduction of the dimensionality. When the number of attributes is large, distances (e.g. euclidean) become less discriminant so that cluster structures cannot be detected (this is one aspect of the curse of dimensionality). Feature selection, for instance through an objective dimensionality reduction, is then compulsory.
    But the use of Principal Components to perform a clustering (unsupervised classification) is not recommended since the components 
with the largest eigenvalues define the axes of maximum variance in the dataset, and these axes are generally not the most discriminative ones to reveal the cluster structure \citep{Chang1983}.

Regarding unsupervised classification, some attempts have been made by K-means
    cluster analysis \citep{ellis2005, Chattopadhyay2007,
    ChattopadhyayGRB2007, Chattopadhyay2008, Chattopadhyay2009b, Chattopadhyay2009a, SanchezAlmeida2010,Fraix2010,Fraix2012,De2016}. Though sophisticated statistical techniques are being developed steadily, unsupervised learning approaches are not widely used across the astronomical community \citep{Fraix-Burnet2015,Souza2017} probably because of the difficulty to relate the multivariate results with physical models. This issue can however be overcome, for instance by comparing statistically the multivariate results with the outcomes of numerical simulations of galaxy and cosmic evolution \citep[e.g.][]{Fraix2012,Genel2014}.
    
    A partitioning of objects into robust groups is possible when the features are independent. 
    Observationally the information is usually summarized into broad-band
    fluxes (magnitudes), slopes (colours), medium-band and line fluxes (Lick indexes) etc, because they can be easily measured and can generally be yielded by models and numerical simulations. However, their relationship to intrinsic physical processes and their mutual influences are most often quite complex. As a consequence, these direct observables cannot suffice to compare two galaxies and the multivariate aspect of the physics of galaxies must be included as well. This is exactly the purpose of the Independent Component Analysis (ICA) that we will use in this paper. 
 
     In this study, we have taken a large dataset from Sloan Digital Sky Survey (SDSS) data archive
    including various observables regarding morphology, chemical composition and
    kinematics and used multivariate statistical techniques to explore and explain
    the underlying diversities. This work presents several novelties for unsupervised classification in astrophysics:
    \begin{itemize}
    \item a large dataset of galaxies
    \item a large number of features
    \item the Independent Component Analysis (ICA)
    \item the adequation of the method (ICA) to the fact that our data are non-Gaussian
\end{itemize}

Some other studies have used the SDSS to get a large dataset \citep[e.g.][]{Cavuoti2014} or use many features \citep[up to 4520!,][]{DIsanto2018} but they perform supervised classification which, in the case of continuous values, is equivalent to a regression problem. The ICA technique has already been used in astrophysics, mainly for source separation \citep{Pires2006,Pike2017,Martins-Filho2018,Sheldon2018} and dimensionality reduction \citep{Richardson2016,Sarro2018} but rarely for unsupervised classification \citep{Mu2007,Das2015}.

     This paper is organized as follows. A brief description of the 
    dataset is given in Section~\ref{data}. The methods are described in Section~\ref{methods}.
    The results and discussion are included in Sections~\ref{results} and \ref{discussion}, respectively. Finally, conclusions are traced in Section~\ref{conclusion}.

\section{Dataset}
\label{data}

The NYU Value-Added Galaxy Catalogue \citep[VAGC][]{Blanton2005,Padmanabhan2008,Abazajian2009} is a cross-matched collection of galaxy catalogues maintained for the study of galaxy formation and evolution\footnote{\url{http://sdss.physics.nyu.edu/vagc/}}. It is based on the Sloan Digital Sky Survey Data Release 7 (SDSS-DR7\footnote{\url{http://classic.sdss.org/dr7/}}).

In the raw table, 2,506,754 objects are available. We have selected only galaxies, by disregarding QSOs and stars using the SDSS flag and subsequently by removing some obviously wrong classification (such as entries with aberrant redshifts or magnitudes). Non-galaxy objects probably remains, but they should show up in the multivariate classification.   We ended up with  865,333 entries. We have then restricted the sample to $z < 0.2$ to avoid too much shift between the wavelength ranges in which the magnitudes are obtained. Finally, only data with a good signal to noise ratio (median S/N per pixel of the whole spectrum $> 10$) were kept. This leaves us with 362,923 galaxies.

We had to limit the number of attributes to keep the computation tractable. For this, we eliminated redundant properties, i.e. features that bear the same information (such as the Sersic profile in different bands), and selected only a few photometric bands and colours. We were careful to keep most of the physical information so that it does not impact much our analysis based on dimensionality reduction through the ICA.
The final set used in our analysis consists of 49
attributes which cover photometry, spectroscopy, morphology, chemical composition and kinematics. Star formation rates and specific star formation rates are also included but not used in the ICA analysis itself since they are not observable features. All these attributes are described in Table~\ref{tab:listparam} and details
are given on the source website\footnote{\url{http://wwwmpa.mpa-garching.mpg.de/SDSS/DR7/raw_data.html}}. Please note that the equivalent width is negative if the line is in emission, \Ha\ and \NII\ are not corrected for possible blending of these two lines, and lines which are absent in spectra are given the value 0.

\section{Statistical Analyses}
\label{methods}

\subsection{Shapiro-Wilk test}

The non-Gaussian nature of the dataset has been explored by the
statistical Shapiro-Wilk test \citep{Shapiro1965} in which the test
statistics is defined by W =
$\sum\limits_{i=1}^{n}a_{i}x_{i}^{2}/\sum\limits_{i=1}^{n}(x_{i}-\bar{x})^{2}$,
where n is the number of observations, $x_{i}$'s are ordered
sample values and $a_{i}$'s are the constants generated from the order
statistics of a sample from normal distribution. In this work a multivariate extension \citep{Alva2009} has been used. The p value of
the test is 2.17$\times10^{-13}$, which is sufficiently small to confidently
reject the null hypothesis. Therefore, the dataset is found to be non-Gaussian in nature.

\subsection{Independent Component Analysis}

We have already mentioned that Principal Component Analysis (PCA)
has been applied by many authors \citep[][etc]{Brosche1973, Whitmore1984, Murtagh1987} but it is not appropriate for
clustering and classification \citep{Chang1983}. Furthermore it is applicable to Gaussian
data which is not the present case. On the other hand, Independent
Component Analysis (ICA) is also a dimension reduction technique, i.e. it reduces the
number of observed features p to a number m (m $<<$ p) of new
variables, the components, but assumes non-Gaussian data \citep{Pfister2018,Hyvarinen1998,Hyvarinen1999a,Hyvarinen1999b}. The second important difference with PCA is that in addition of being uncorrelated, the components are also mutually independent \citep[for  details  regarding comparison between PCA and ICA see Section 3 of ][ and references therein]{Chattopadhyay2013}.

Mathematically speaking, let $X_{1}$, $X_{2}$, $X_{3}$, ...,
$X_{p}$ be p random vectors (here p features, p = 47) and n
(here 362,923) be the number of observations of each $X_{i}$,
(i = 1, 2, 3, ..., p).

Let X = AS, where S = $[S_{1}, S_{2}, S_{3}, ..., S_{p}]^{'}$ is a
random vector of hidden components $S_{i}$, (i = 1, 2, 3, ..., p)
such that $S_{i}$'s are mutually independent and A is a non
singular matrix. Then the objective of ICA is to find S by inverting A,
i.e., S = $A^{-1}$X = WX. Since ICA is able to separate independent components (sources) present in a signal, W is called the unmixing
matrix \citep[for more details see][ and references
therein]{Comon1994,Chattopadhyay2013}. Independence is obtained by maximizing the non-Gaussianity using negentropy.

Presently there is no good method available for the determination of
the optimum number of ICs. In this work, the optimum number of ICs
have been chosen by the optimum number of Principal Components (PCs)
\citep{Albazzaz2004, Chattopadhyay2013, Eloyan2013}, to find m (m $<<$ p) \citep{Chattopadhyay2007,Babu2009,Fraix2010,
Chattopadhyay2010, Chattopadhyay2013}. We have
first performed PCA to find the significant number of ICs. In PCA
the maximum variation with significantly high eigenvalue (viz.
$\lambda$ $\sim$ 1) was found to be almost 90\% for nine PCs.
Hence, we have chosen nine ICs  (i.e. here m = 9) for cluster analysis (CA). Here it is worth mentioning that the choice of the ICs is a difficult question \citep[see][]{Kairov2017} but we have checked that the set of most correlated original features (Section~\ref{res:corrIC} and Table~\ref{tab:ICparam}) is stable through multiple runs of the algorithm. Note that for our dataset PCA takes 132 seconds whereas ICA takes less than 1 minute (Intel(R) Core TM i3-2100 CPU @ 3.10GHz, RAM: 2.00 GB, Windows 10, 64 bit operating system) and we have used the library functions publicly available for ICA and PCA in R.

\subsection{K-means cluster analysis}

K-means cluster analysis (CA) is a multivariate technique for finding coherent
groups in a dataset giving information of the underlying
structure. In this method:  \begin{itemize}
    \item Each object must belong to a single cluster.
    \item Each cluster must contain at least one object.
    \item All the objects are distributed among K clusters by satisfying the first two properties.
\end{itemize}
 Details of algorithm and applications are found
in \citet{kmeans1967, Chattopadhyay2009a, Chattopadhyay2010, Chattopadhyay2012, Chattopadhyay2013,Das2015}.

The number K of groups is an input to the algorithm. The optimum value of K is found as follows. For a particular choice of K initially we compute a distortion measure $d_K$, given by,  $d_K = (1/p)
min _x E[(x_K - c_K)^{'} (x_K - c_K)]$ which is the distance of $x_K$ vector (data point) from the centroid $c_K$ of the corresponding group. Then we compute the value of jump $J_K = (d_K^{-p/2} - d_{K-1}^{-p/2}$) where p is the total number of variables (here total number of independent components under consideration i.e. p = 9). The above $d_K$ and $J_K$ values are computed for several choices of K starting from K = 1, 2, 3, etc. The maximum value of the curve obtained by plotting $J_K$ versus K gives the optimum value of K  \citep{Sugar2003}. 

In this study, we have performed a K-means CA with respect to the ICs and have
found the optimum number of groups to be K = 10. We name the groups K1 to K10.

\section{Results}
\label{results}

\subsection{Properties of the ICs}
\label{res:corrIC}

\begin{table}
\centering \caption{Observed attributes with highest correlation coefficients (in parentheses) for each Independent Component.}
\label{tab:ICparam}
\begin{tabular}{ll}
\hline \hline IC & Most influencial attributes \\ \hline
ICA1 & \OII (0.83), Lick\_Fe5015 (0.53) \\
ICA2 & \vdisp\ (0.97), J (-0.7) \\
ICA3 & \sigmabalmer\ (-0.60) \\
ICA4 & \OIIIfive\ (0.98) \\
ICA5 & Lick\_Fe5015 (0.83) \\
ICA6 & \Sersicrninety (-0.99) \\
ICA7 & \sigmaforb\ (-0.60), \sigmabalmer\ (0.42) \\
ICA8 & \Sersicamp\ (-0.99) \\
ICA9 & \NII (0.79), \Ha (0.75), \\
     & \SII (0.72) \\
\hline
\end{tabular}
\end{table}

Performing Independent Component Analysis for the VAGC dataset,  we have taken nine significant ICs. Then we have computed the correlation coefficients of each component with the attributes to understand their physical meaning. We list only the most influencial attributes in Table~\ref{tab:ICparam}. 

From Table~\ref{tab:ICparam}, it is clear that the nine ICs represent five
kinds of properties: 1) velocity dispersion (ICA 2, ICA 3, ICA 7),
2) ionisation (ICA 4, ICA 9), 3) metallicity (ICA 1, ICA 5), 4)
surface brightness (ICA 8) and 5) structural properties (ICA 6).
This shows that the ICA successfully retrieve the main physical ingredients of galaxies, while also taking into account their interactions through the linear combinations of the observed features. 

This result is indeed very significant. Usually, a given physical process (star formation, nuclear activity, kinematics, radiation characteristics, etc) is investigated through an obvious observable feature, leading by the way to the traditional classifications mentioned in Section~\ref{introduction}. Here, each of the ICs is dominated by some features specific to a particular physical property of galaxies, without any a priori selection. In addition, each IC is not limited to the dominant features, it also includes the other ones with some lower weights thus taking into account the complex interplay between observed attributes.

The absence of some traditional features such as colour, line ratios or morphology among the dominant features is striking and will be discussed in Sections~\ref{res:colmag}, \ref{res:BPT} and \ref{res:morphology} respectively.

These nine independent components are used instead of the initial 47
attributes for Cluster Analysis, very substantially reducing the 
dimensionality of the dataset while keeping all the physical information.

\subsection{Properties of the galaxies in the ten groups}

\begin{table}
\centering \caption{Distribution of galaxies in the ten groups
found with K-means.}
\label{tab:distribgroups}
\begin{tabular}{lrr}
\hline \hline
Group & Number of galaxies & Percentage \\
\hline
K1 & 1,375 & 0.38 \\
K2 & 16,325 & 4.50 \\
K3 & 68,050 & 18.75\\
K4 & 109,188 & 30.08 \\
K5 & 79,576 & 21.93 \\
K6 & 17,336 & 4.78 \\
K7 & 7,846 & 2.16 \\
K8 & 40,045 & 11.03 \\
K9 & 10,552 & 2.91 \\
K10 & 12,630 & 3.48
\\ \hline
\end{tabular}
\end{table}

The cluster analysis divided the galaxies into ten groups, K1-K10. The distribution of galaxies within these groups is given in Table~\ref{tab:distribgroups}.
It appears that the four groups K4, K5, K3, K8,  in decreasing importance, already gather
82\% of the objects (52\% with K4 and K5 only), the K1 group being very small. 

Figure~\ref{fig:boxplotsIC} shows the distribution of the ten groups in the nine ICs. IC1 and IC3 do not seem very discriminant, but all the others explain the specificities found by the CA algorithm. Many dimension reduction techniques creates new variables (here the IC components) which are linear or non-linear combinations of the original features. They are a mathematical representation of the data that is not intended to have a physical meaning. This is a mathematical space in which the data appear structured in groups. The physical interpretation should aim at understanding these groups in the physical space, not in the IC space. It should be clear that traditional approaches fail since each group is characterized by a complex multivariate physics. It is certainly simpler to summarize galaxies as blue, star forming, or spiral, but this is obviously very limited. The solution is thus to have more complex descriptions of galaxy classes, and use models and numerical simulations to interpret the results of multivariate analyses. 

The boxplots shown in Figure~\ref{fig:boxplots} summarize the statistics for the ten groups of the 47 attributes used for the clustering analysis plus SFR, specSFR, Sersic\_r50\_R and the redshift. The order and the names of the groups given in Table~\ref{tab:distribgroups} is arbitrary and has been chosen to smoothen the variation of the boxplots from group K1 to K10 for most observed features. This nomenclature thus bears absolutely no physical or temporal evolutionary relationships between these groups.

It is interesting to note that the dispersions are nearly always relatively small, indicating that the groups found by the cluster analysis are quite homogeneous. This is particularly striking for the biggest groups, K4 and K5. There are often large overlaps, but also clearly separated properties distributions between groups in many instances. For some properties, such as \Haabs, J-K and H-K, there is very little variation from group to group.

The small group K1 stands out in many features, especially in \OIIIfive. This group is however close to the groups K2 and K3 for many properties, such as \CaK, \Sersicn\ or Lick\_G4300. With K4 they share a low velocity dispersion with a clearcut split with respect to the other groups K5 to K10. This split is also visible for \Sersicn\ and also somehow on the redshift diagram.

The specific SFR shows a remarkably regular decrease from K1 to K10, while D4000$_\mathrm{n}$ has the opposite behaviour.

Groups K6 and K7 gather galaxies which are far much larger than all the other ones. These two groups are not so different in the other observable features, except may be that K7 has a higher average redshift, pointing to possible measurement errors. It is important to recall that most attributes are determined automatically, so aberrant values may appear. However, these two groups are not very small, making such errors (hopefully) improbable. Also, such large sizes are not uncommon \citep[e.g.][]{Guo2009}. Further analyses of the galaxies of these groups, for instance using images, should be done before any physical interpretation. This is beyond the scope of this first paper, we simply emphasize that the CA analysis is able to separate these objects into distinct groups.

\subsection{Colour-magnitude diagram}
\label{res:colmag}

\begin{figure}
\centering
\includegraphics[width=\columnwidth]{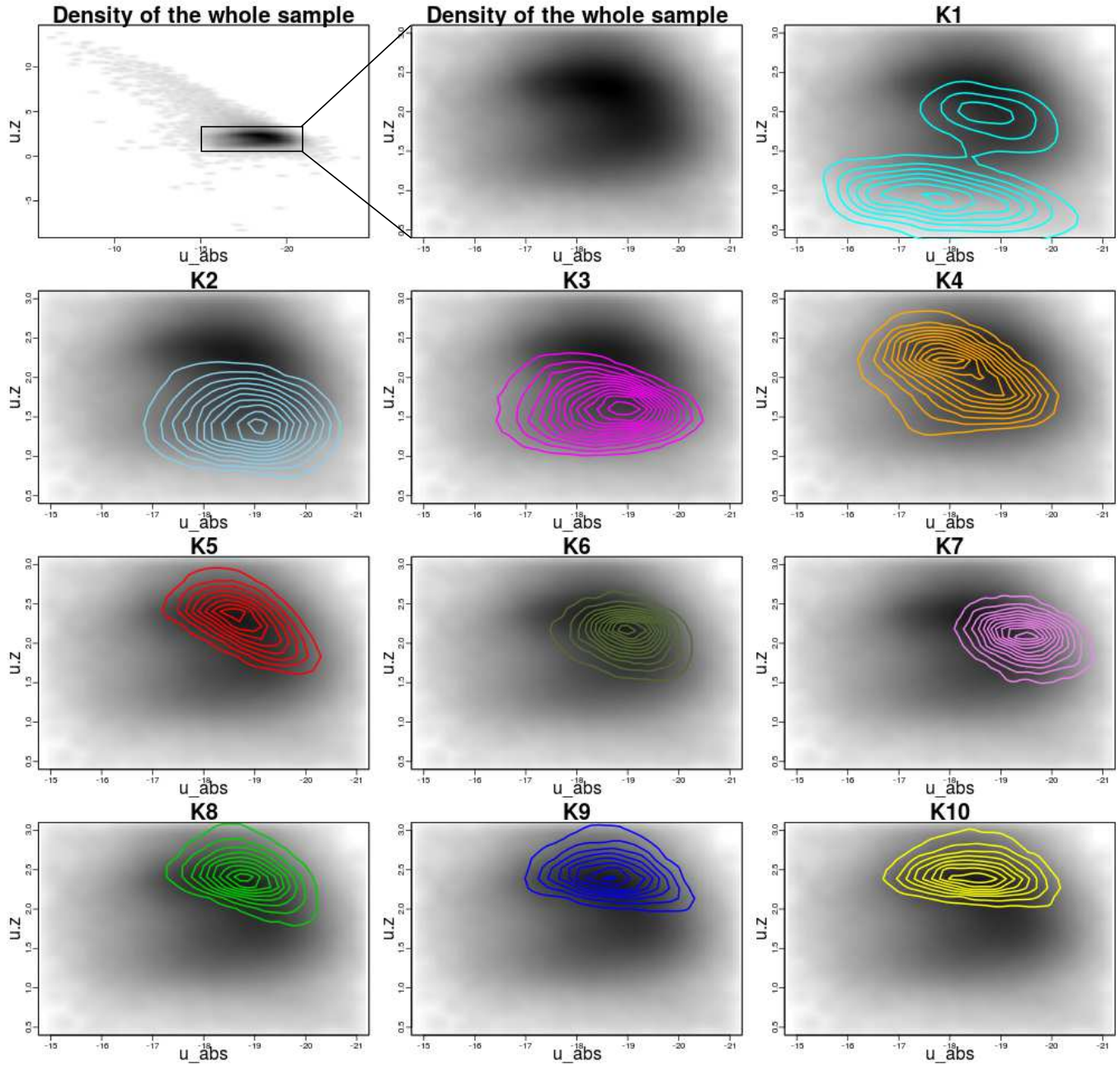}
\caption{Colour-magnitude (u-z vs. U) diagrams of the whole sample and for each of the groups.}
\label{fig:colmag}
\end{figure}

The well known bimodality of galaxies is seen on the colour-magnitude diagrams in Fig.~\ref{fig:colmag} with a crescent shape of the distribution of the whole sample, with the so-called red and blue branches (or sequences). 

The groups K5, K8, K9 and K10 clearly belong to the red branch, while K2 to K3 are essentially on the blue one. The group K4 spans both branches, but peaks mainly on the red branch and includes the region in between often called the green valley. Interestingly, the groups K6  and K7 that have the largest galaxies in our sample, are both part of this green valley.
The very small group K1 appears peculiar, with a very blue part, and 25\% of its galaxies belonging to the red branch. 

The correspondence between our groups and the rough usual division in red, green or blue regions of the plot is quite good despite the fact that no colour nor magnitude are involved in the independent components used for the classification. This is very likely due to a certain degree of correlation among some properties of galaxies, which multivariate clustering, by determining more subtle and objective categories, is able to stress.

Some groups extend to several regions, such as the big group K4. However, the bivariate colour-magnitude diagram alone cannot reasonably encompass all the physics of galaxies. In this respect, it is interesting to compare our groups with the Figure~2 of \citet{Alatalo2014} on which some properties, like SFR or \Hd, are plotted. Clearly these properties do not peak and are not confined to only one of the blue, red or green regions, similarly to our groups.

\subsection{Emission-line diagnostic diagrams}
\label{res:BPT}

\begin{figure}
\centering
\includegraphics[width=0.7\columnwidth]{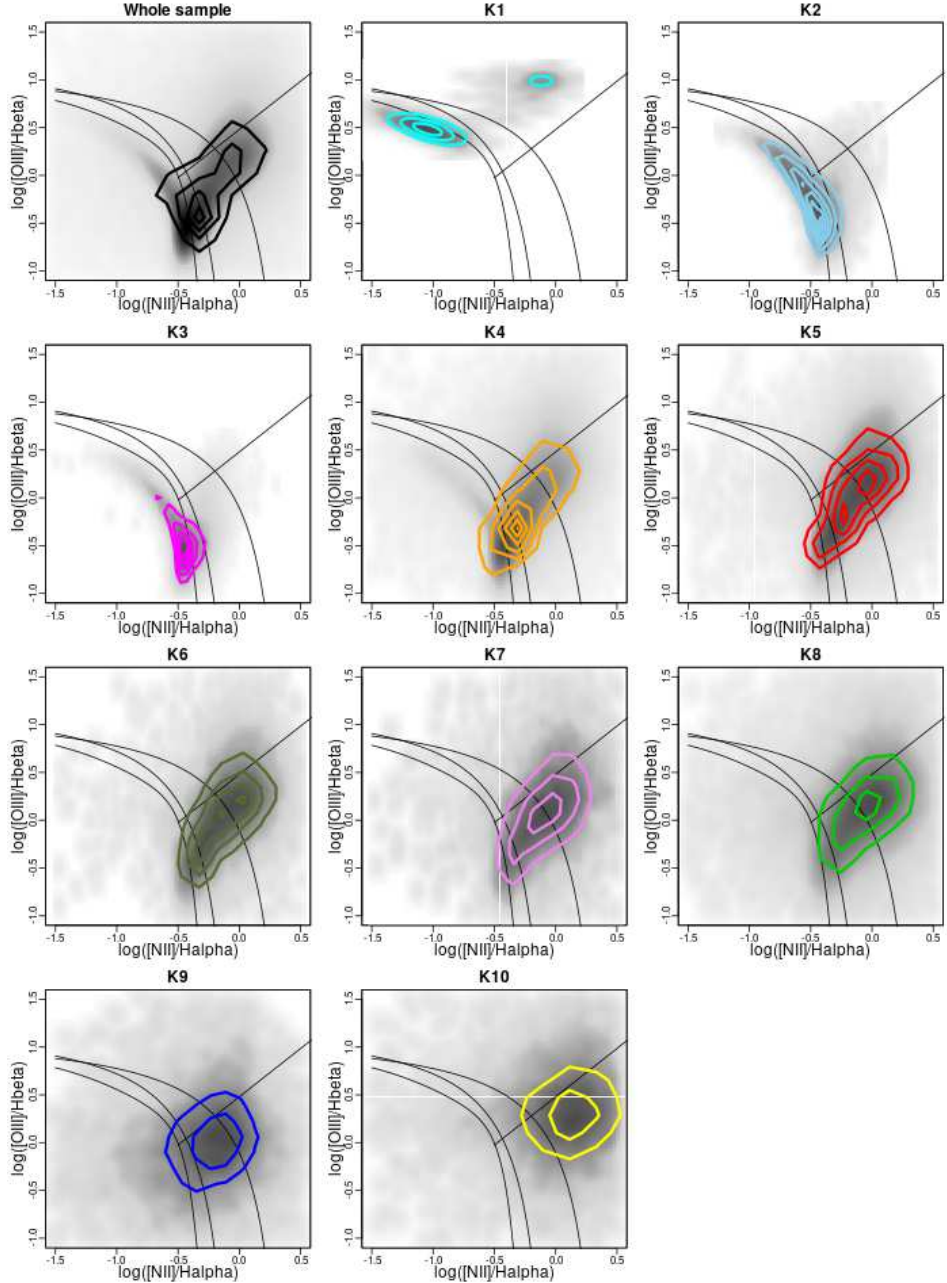}
\caption{The distribution of galaxies in the groups K1 to K10 in
the BPT diagram: [OIII]/H$\alpha$ vs [NII]/H$\beta$. Curves are from \citet{Kewley2001,Kauffmann2003a,Stasinska2006} (right to left respectively). The bottom left zone is for star forming regions, the upper zone is for AGNs and Seyfert galaxies, and the bottom right zone is for LINERs.}
\label{fig:BPT}
\end{figure}

Two emission line ratios were recommended by \citet[][ hereafter BPT]{Baldwin1981} and are often used to discriminate between star forming and AGN-dominated galaxies. This diagram is also the basis for more refined empirical classifications based on theoretical population synthesis and photoionization models \citep[][]{Veilleux1987,Kauffmann2003a,Kewley2006,Kewley2013a}. Two other similar diagrams, BPT-SII and BPT-OI, were proposed by \citet{Veilleux1987}. They will be presented in subsequent papers.

The standard classification scheme on this diagram is based on equations of curves separating the different classes. These cuts are sharp, somewhat arbitrary. Its main goal is to distinguish between different ionization source (basically thermal or non-thermal) while the properties used for our clustering analysis are not limited to this peculiar aspect of galaxies. It also appears that the classification based on these diagrams is not clearcut. For instance, "the current LINER classification scheme encompasses two or more types of galaxies, or galaxies at different stages in evolution" \citep{Kewley2006}. \citet{CidFernandes2010} proposed a new cut between Seyfert and LINERs to resolve this ambiguous class.  They also present an interesting and critical discussion of the classifications based on the BPT diagrams. An important reminder in their discussion is that the star-forming region delimitation is rather arbitrary in the lower part of the diagram where most of the galaxies lie. Also, the upper right part of the diagram is not composed of pure AGNs. Finally, different cuts are proposed by different authors \citep[e.g. ][]{Kewley2001,Lamareille2010}. Interestingly, all these works tend to suggest that more parameters are probably required to fully understand different kinds of galaxies and ionization processes \citep{Richardson2016}.

Indeed, an objective classification with soft frontiers performed with the diagnostics line ratios indicates a somewhat different picture with only four categories, some including for instance sub-divisions like strong or weak AGNs \citep{Souza2017}. However while being multivariate, this study is only three-dimensional. 

An additional limitation of the classifications based on single kinds of ionization processes is that the galaxies are indeed very probably a mixture of different regions \citep{Belfiore2016}.

These limitations on the significance of the cuts should be kept in mind when comparing our multivariate clustering results and diagnostics diagrams (Fig.~\ref{fig:BPT}). We see that K2 and K3 have no AGN, hence are pure star forming galaxies,
whereas the very small group K1 is quite peculiar with two peaks, one in the pure star forming region, the other one in the pure AGN zone.   Groups K4 and K5 are intermediate between star forming galaxies and AGNs. 
All the other groups (K6, K7, K8, K9 and K10) are LINERs (Fig.~\ref{fig:BPT}).

Our result is in the line of the discussion found in the literature as presented above in the sense that our groups can be clearly placed in the different zones of the BPT diagram with some fuzziness. This agreement is remarkable because no line ratios are involved in the ICs used for our classification, only H$_{\alpha}$, [OIII] and [NII] but not H$\beta$.

\subsection{Morphology}
\label{res:morphology}

\begin{figure}
\centering
\includegraphics[width=\columnwidth]{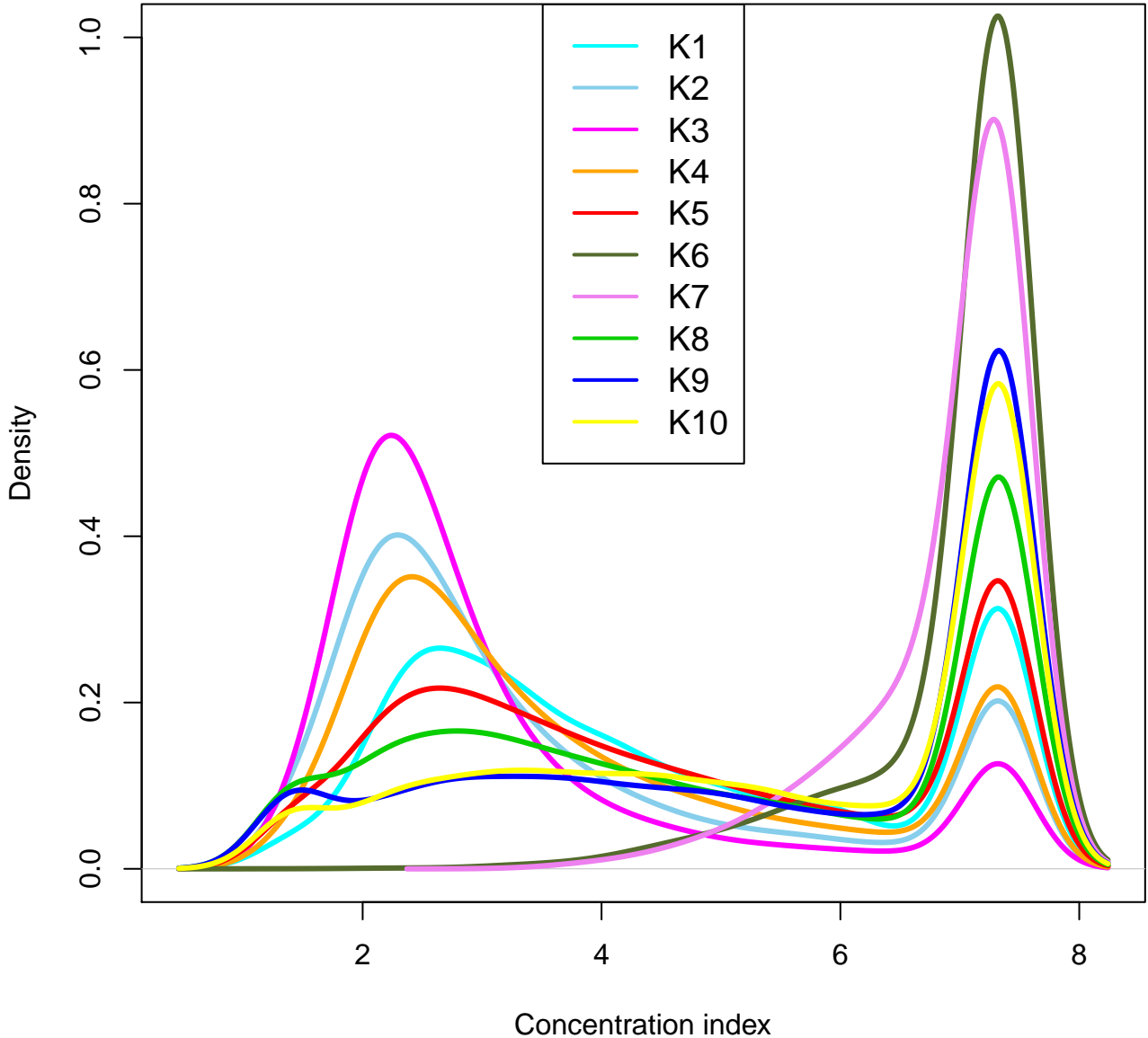}
\caption{Density distribution of the concentration index  \SersicC\ of galaxies in the groups for K5 to K7.}
\label{fig:concentration}
\end{figure}

 In Fig.~\ref{fig:concentration}, we plot the density distribution of the concentration index \SersicC\ = \Sersicrninety\ / \Sersicrfifty\ for all the
ten groups. Late-type morphologies are characterized by a low \SersicC\  \citep[$< 2.6$, ][]{Strateva2001}. Note the strong dichotomy with two peaks at \SersicC $\simeq 2.5$ and $7.5$.

The groups K6 and K7 are nearly entirely made of early-type galaxies while all the other groups are mixtures of both categories, with a higher fraction of late-type galaxies in K2 and K3 and of early-type ones in K5, K10, K8, K9.

The dichotomy between early- and late-type galaxies is thus roughly recovered despite the fact that no morphological indicator such as the concentration index and \Sersicn\ are present in the ICs. This absence probably explains why some of our groups have a mixture of both types, but it is also an indication that the morphology is not very discriminative when the whole multivariate physics of galaxies is considered.

\section{Discussion}
\label{discussion}

In this first paper, we intend to focus mainly on the technique used for the multivariate classification of our large sample, and to provide a rather general description of the physical properties of the ten groups we have found to demonstrate that the classification obtained from the IC components is physically meaningful. Naturally, the multivariate grouping seems more complicated than traditional classifications, so that more detailed studies on some specific aspects, such as nucleus activity or SFR, will be made in subsequent papers. For now, from the results described in the previous section, we can distinguish four categories of groups that mainly match usual classes of galaxies.

\subsection{Groups K1, K2, K3}

 The three groups K1, K2, K3 belong to the so-called blue sequence and are active sites of star formation as supported by high
\Hbabs, \Hdabs, \Hgabs, specSFR, low metallicity and high \OIIIfive.

They are also the less massive objects of the sample, especially K1. They have higher values of \Hdabs\  thus containing young stellar populations but the widths of its distribution are
not similar in these groups. There is maximum scatter in K1 and
minimum scatter in K2. They all have a high SFR, but the very small group K1 has a particularly high specSFR. The large scatter in \OIIIfive\ in the latter group indicates that the galaxies belonging to this K1 group contain gas at high temperature to form new generation of stars.
Hence K1 has a stellar population which is a mixture of various ages as a result of star bursts of recent origin.

Though K2 and K3 are dominated by late type galaxies, they also have small fractions of
spheroidals. The galaxies in these groups have a low \Sersicn\ lying between 0 and
2  which indicates that the bulges in the spheroidals are not also very
pronounced, i.e., these galaxies may be in a formation stage.

\subsection{Groups K4, K5}

The two K4 and K5 groups gather about half of our sample. From the
colour-magnitude diagram (viz. Fig.~\ref{fig:colmag}), it is clear that they
 are mainly on the red branch with some extensions towards the green and blue regions, especially for K4. However, their \Sersicn\  indexes lie between 2 - 4 which show that some of the galaxies are young and in the formative stages of their bulges and some of them
have well developed bulges. The SFR and metallicities in these
galaxies are intermediate between highest and lowest values. The two groups
differ largely in their light element abundances: K5 has
larger value than K4. Regarding \Hdabs\ 
the values lie between 0.5 and 2.5 with pronounced scatter. 
Groups K4 and K5 have globally intermediate values of SFR and more particularly of specSFR with respect to the other groups. Most of the properties like
metallicity, SFR, \Hdabs, specSFR, D4000$_\mathrm{n}$ and \SersicC\  have large scatters which indicate
that K4 and K5 have a certain mixture of old and young populations, as well as
of active and passive galaxies with well defined bulges for K5 and more pseudo bulges for K4.

Galaxies in the K5 group have higher velocity dispersion hence are more massive than those belonging the group K4. The group K5 is more abundant in helium
enriched population, which is a signature of second
generation stars. So K5 has slightly older population than K4.
This is also evident from the D4000$_\mathrm{n}$ peaks: K5 has a larger peak
at higher D4000$_\mathrm{n}$ whereas the opposite is true for K4. Also $C$ values show that the two groups have populations of late type spirals, but in much higher proportion in K4.

As a conclusion, despite the fact that K4 and K5 galaxies mainly belong to the red sequence, and would consequently be considered as quenched, they have somewhat average values for most features, and possess some star forming objects. 

This result may seem inconsistent, but this is the outcome of a multivariate analysis that gathers galaxies sharing not merely a few properties, but many. However, we cannot exclude that this apparent inconsistency might be due to two possible limitations of our study.
The first one is that the clustering algorithm used (K-means) may not be the most suitable technique for the sample, and in particular for the subsamples made with K4 and K5 galaxies. The second explanation might be that even with the information contained in the 47 attributes used in our study and summarized in the 9 ICs used for the classification, it is not possible to distinguish sub-classes within these two big groups. This may be due to several reasons: lack of more distinctive information, uncertainties that smear out differences among variable values of different groups, or the fact that the observables used here are  integrated values over an entire galaxy probably mixing up several different regions within these big and complex systems. Further investigations will be made in another paper.

\subsection{Groups K6, K7}

Groups K6 and K7 are globally similar with mostly average values. They are often close to groups K4 and K5, but they have remarkable high $C$ together with very high \Sersicrfifty\ and \Sersicrninety. Hence they are made up of very big and large galaxies, and thus spheroidals. They occupy the same region in  the colour-magnitude diagram, essentially corresponding to upper part of the so-called green valley. They have a relatively old population as based on D4000$_\mathrm{n}$, but with a small fraction of younger stellar populations and some star formation ongoing. 

These galaxies thus appear to be in a transitional phase of being quenched but they are really big and massive galaxies which have no equivalent in the red branch members in our sample (red branch that is usually considered as the end fate of all galaxies). The groups K6 and K7 may thus represent the dead end of big galaxies.

\subsection{Groups K8, K9, K10}

The three groups K8, K9 and K10 are rather specific: they have the lowest specSFR, high velocity dispersion, high metallicity, high Sersic indexes and C $>> 2.6$ for most of their members. All this indicates that they are early type spheroidal galaxies and quenched. They are more concentrated at the center and massive in nature. They have a high abundance of oxygen (viz. Fig.~\ref{fig:boxplots}, \OIIIfour, especially in K10) which might be due to the explosion of massive supernovae (Pop II objects). They have low \Hdabs\ values with small scatters and high D4000$_\mathrm{n}$ values. They have well developed bulges as seen from their high \Sersicn\  values ($\sim$ 4, Fig.~\ref{fig:boxplots}). The forbidden line are pronounced in K10, indicating that these galaxies have enough neutral gas.

\subsection{Independent components}

We have seen that the nine ICs used to perform the cluster analysis represent a complete physical description of galaxies: velocity dispersion, ionisation, metallicity, surface brightness and structure. It is important to realize that they have not been selected on this basis. It should also be clear that using the nine ICs describing five properties is not equivalent to selecting five observables representing these properties, or even nine to match the number of ICs, since the latter are linear combinations of the observable features that reveal some latent multivariate relationships.

The most remarkable outcome is that the groups built from the ICs very satisfactorily match the traditional categorization of galaxies, despite the fact that none of the traditional observables is involved in the nine components. We might see this correspondence from two points of view: either the multivariate analysis is unnecessary, or, as we tend to think, the traditional approach is only an approximation of the complexity of galaxies and is thus incomplete. We obtain more groups than the usual classification, but this is because we are not able to see into a high-dimension space.

Among the observable features involved in the ICs, it is striking to see that the equivalent widths of several lines are important. Also the velocity dispersion and the line widths take their share. The size (radius) and mass (with the J magnitude) seem to matter. On the contrary, the absence of any colour, of the concentration index, related to the global morphology, or even the index D4000$_n$, is remarkable. 

\section{Summary and Conclusion}
\label{conclusion}

In this study, we have classified a large dataset of
galaxies with a large number (47) of morphological, photometric and
spectroscopic parameters compiled from VAGC/SDSS data archive. We have
used two sophisticated statistical methods. Firstly we have performed a dimension reduction using the  ICA analysis  which is
appropriate for a non-Gaussian dataset such as this one. Secondly, k-means cluster analysis has been used to find structures in the parameter space defined by nine Independent Components to obtain ten coherent groups.

The most correlated observable features involved in the nine ICs represent a complete physical description of galaxies (velocity dispersion, ionisation, metallicity, surface brightness and structure). Despite neither morphology nor colour nor emission-line ratios are  among the most influential features in the ICs, our ten groups can be essentially identified as early- or late-type dominated and placed in the colour-magnitude and the line diagnostic (BPT) diagrams in agreement with the literature.

However, our classification is not empirical and is based on a rather complete physical description of the galaxies. The ten groups then allow for a more refined interpretation of galaxy diversity.

Recovering known divisions of galaxies gives us confidence in our unsupervised multivariate clustering analysis, and subsequent papers will be devoted to the astrophysical interpretation and implications of each of the groups we have identified. In particular, the mixture of traditional populations in some of our groups must be understood in the multivariate frame, compared to models and numerical simulations of galaxies that only can provide a multivariate physical understanding, without forgetting the context of evolution.

\section*{Acknowledgements}

We thank Fabrice Lamareille and Emmanuel Davoust for having compiled for us the sample used in this analysis. We warmly thank Malgorzata Siudek for invaluable discussions. TK and DFB are thankful to the Indo-French Centre for Applied Mathematics (Bangalore, India) for providing partial support.
We thank the two referees for their detailed and very constructive reports. 

 Funding for the Sloan Digital Sky Survey (SDSS) has been provided by the Alfred P. Sloan Foundation, the Participating Institutions, the National Aeronautics and Space Administration, the National Science Foundation, the U.S. Department of Energy, the Japanese Monbukagakusho, and the Max Planck Society. The SDSS Web site is http://www.sdss.org/.

The SDSS is managed by the Astrophysical Research Consortium (ARC) for the Participating Institutions. The Participating Institutions are The University of Chicago, Fermilab, the Institute for Advanced Study, the Japan Participation Group, The Johns Hopkins University, Los Alamos National Laboratory, the Max-Planck-Institute for Astronomy (MPIA), the Max-Planck-Institute for Astrophysics (MPA), New Mexico State University, University of Pittsburgh, Princeton University, the United States Naval Observatory, and the University of Washington.

\bibliographystyle{aasjournal}
\bibliography{vagc} 

\clearpage

\appendix
\counterwithin{figure}{section}
\counterwithin{table}{section}
\renewcommand\thefigure{\thesection\arabic{figure}}   
\renewcommand\thetable{\thesection\arabic{table}}

\section{Complementary tables and figures}

\newpage

\begin{table*}
\centering
\caption{All the parameters of the dataset used for the analysis.\footnote{See
\url{http://wwwmpa.mpa-garching.mpg.de/SDSS/DR7/SDSS\_line.html} for
more details.}}
\label{tab:listparam}
{\renewcommand{\arraystretch}{0.5}

{\footnotesize
\begin{tabular}{ll}
\hline Parameter & Description \\ \hline
\vdisp & Estimated velocity dispersion from spectrum \\
$u$ & u absolute magnitude (log of intensity) \\
$J$ & $J$ absolute magnitude \\
\Sersicamp & The best fit to the variable ``A" in band R (nanomaggies\footnote{The flux f in nanomaggies is defined in \citet{Blanton2005} as magnitude$=22.5-5\log10(f)$.}/$arcsec^{2}$): \\ 
& describes the radial distribution of light \\
\Sersicn & The best fit to the Sersic index ``n" in band R \\
sigma\_balmer & Velocity dispersion ($\sigma$ not FWHM) measured simultaneously in all \\
& the Balmer lines in km/s \\
sigma\_forb & Velocity dispersion ($\sigma$ not FWHM) measured simultaneously in all \\
& the forbidden lines in km/s \\
oii\_3729\_seqw & The equivalent width of the continuum-subtracted emission line \\
&with the other emission lines subtracted off \\
neiii\_3869\_seqw & \hskip 5 cm " \\
oiii\_4363\_seqw (\OIIIfour) & \hskip 5 cm " \\
oiii\_5007\_seqw (\OIIIfive) & \hskip 5 cm " \\
hei\_5876\_seqw & \hskip 5 cm " \\
oi\_6300\_seqw & \hskip 5 cm " \\
h\_alpha\_seqw (\Ha)& \hskip 5 cm " \\
nii\_6584\_seqw & \hskip 5 cm " \\
sii\_6731\_seqw & \hskip 5 cm " \\
\Hdabs & Equivalent width for absorption lines \\
\Hgabs & \hskip 1.6 cm " \\
\Hbabs & \hskip 1.6 cm " \\
\Haabs & \hskip 1.6 cm " \\
\CaK & \hskip 1.6 cm " \\
Ca\_h\_abs & \hskip 1.6 cm " \\
Na\_d\_abs & \hskip 1.6 cm " \\
Lick\_CN2 & Stellar absorption line (Lick) index \\
Lick\_Ca4227 & \hskip 0.7 cm " \\
Lick\_G4300 & \hskip 0.7 cm " \\
Lick\_Fe4383 & \hskip 0.7 cm " \\
Ca4455 & \hskip 0.7 cm " \\
Lick\_Fe4531 & \hskip 0.7 cm " \\
Lick\_c4668 & \hskip 0.7 cm " \\
Lick\_Hb & \hskip 0.7 cm " \\
Lick\_Fe5015 & \hskip 0.7 cm " \\
Lick\_Mgb & \hskip 0.7 cm " \\
Lick\_Fe5270 & \hskip 0.7 cm " \\
Lick\_Fe5335 & \hskip 0.7 cm " \\
Lick\_Fe5406 & \hskip 0.7 cm " \\
Lick\_Fe5709 & \hskip 0.7 cm " \\
Lick\_Fe5782 & \hskip 0.7 cm " \\
Lick\_NaD & \hskip 0.7 cm " \\
Lick\_Hd\_A & \hskip 0.7 cm " \\
D4000$_\mathrm{n}$ & The break in the spectrum at 4000 \AA \\
SFR & Star Formation Rate \\
specSFR & Specific Star Formation Rate \\
\Sersicrzero & The best fit to the variable r\_0 in R band (arcsec) \\
\Sersicrninety & 90\% light radius of best fit model in R band (arcsec) \\
\SersicC & Concentration Index: ratio between \Sersicrninety\ and \Sersicrfifty  \\
$u-z$ & Magnitude u minus magnitude z (u-z) \\
$J-H$ &  \\
$H-K$ &  \\
\hline
\end{tabular}
}
}
\end{table*}

\clearpage


\begin{figure*}
\setcounter{figure}{0}  
\centering
\includegraphics[width=0.9\linewidth]{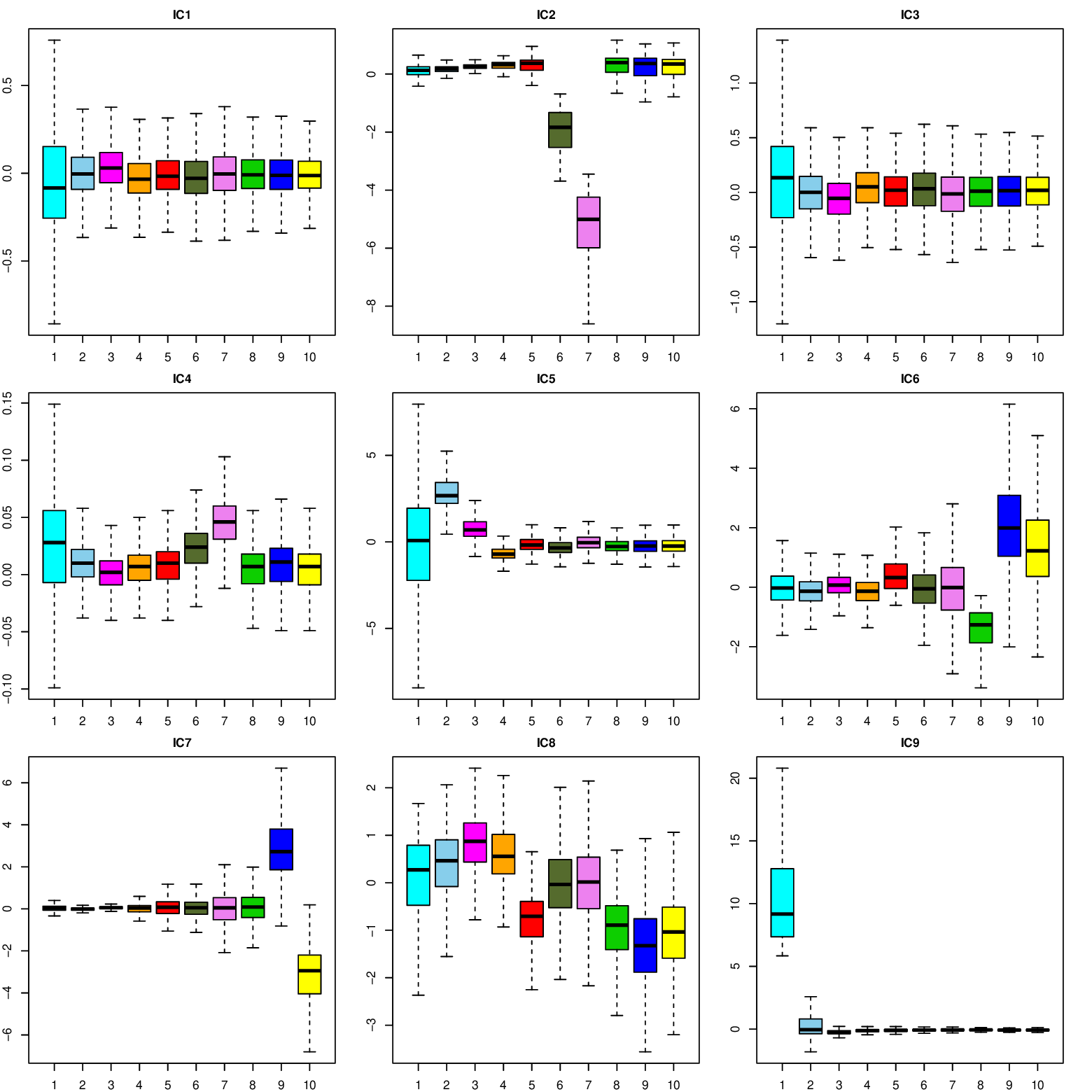}

\caption{Boxplots for the IC components.} \label{fig:boxplotsIC}
\end{figure*}

\begin{figure*}
\centering
\includegraphics[width=0.75\linewidth]{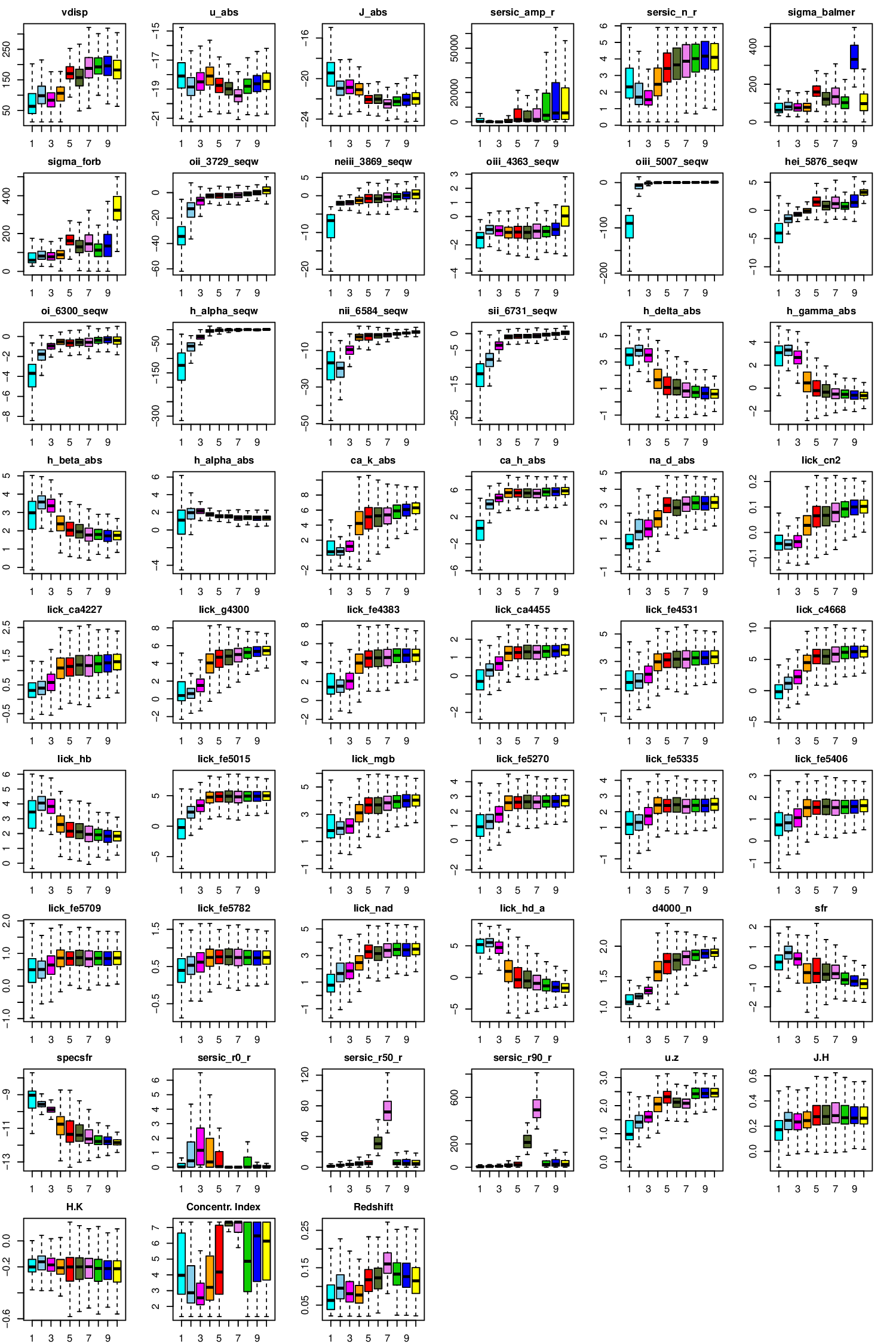}
\caption{Boxplots for the 47 attributes used for the analysis, plus SFR, specSFR, Sersic\_r50\_R and the redshift.} \label{fig:boxplots}
\end{figure*}

\end{document}